\documentclass[aps,prb,twocolumn,preprintnumbers,amsmath,amssymb,superscriptaddress]{revtex4}%

\usepackage{graphicx}%
\usepackage{dcolumn}
\usepackage{amsmath}
\usepackage{color}
\usepackage{multirow}

\makeatletter
\def\btt#1{\texttt{\@backslashchar#1}}%
\DeclareRobustCommand\bblash{\btt{\@backslashchar}}%
\makeatother

\begin{document}

%\preprint{PREPRINT (\today)}

\title{Isotope effect on the transition temperature $T_c$ in Fe-based superconductors: \\ the current status}
\author{Rustem~Khasanov}
 \email[Corresponding author: ]{rustem.khasanov@psi.ch}
 \affiliation{Laboratory for Muon Spin Spectroscopy, Paul Scherrer
Institute, CH-5232 Villigen PSI, Switzerland}
%
%\author{R.~Khasanov}
% \email[Corresponding author: ]{rustem.khasanov@psi.ch}
% \affiliation{Laboratory for Muon Spin Spectroscopy, Paul Scherrer
%Institute, CH-5232 Villigen PSI, Switzerland}
%
%
%\author{H.~Keller}
%\affiliation{Physik-Institut der Universit\"{a}t Z\"{u}rich,
%Winterthurerstrasse 190, CH-8057 Z\"urich, Switzerland}

\begin{abstract}
The results of the Fe isotope effect (Fe-IE) on the transition temperature $T_c$ obtained up to date in various Fe-based high temperature superconductors are summarized and reanalyzed by following the approach developed in [Phys. Rev. B {\bf 82}, 212505 (2010)]. It is demonstrated that the very controversial results for Fe-IE on $T_c$ are caused by small structural changes occurring simultaneously with the Fe isotope exchange. The
Fe-IE exponent on $T_c$ [$\alpha_{\rm Fe}=-(\Delta T_c/T_c)/(\Delta M/M)$, $M$ is the isotope mass] needs to be decomposed into two components with the one related to the structural changes ($\alpha_{\rm Fe}^{\rm str}$)
and the genuine (intrinsic  one, $\alpha_{\rm Fe}^{\rm int}$).
The validity of such decomposition is further confirmed by the fact that $\alpha_{\rm Fe}^{\rm int}$ coincides
with the Fe-IE exponent on the characteristic phonon frequencies $\alpha_{\rm Fe}^{\rm ph}$  as is reported in recent EXAFS and Raman experiments.
\end{abstract}

\maketitle

\section{Introduction}

Historically, the isotope effect plays  an important role in elucidating the origin of the pairing interaction leading to the
occurrence of superconductivity.
The discovery of the isotope effect on the superconducting transition temperature $T_c$ in metallic Hg \cite{Maxwell50,Reynolds50} in the year 1950 provided the key experimental evidence for phonon-mediated pairing and leads to the subsequent formulation of the  BCS theory.

An involvement of the lattice degrees of freedom into the supercarrier formation is generally considered by measuring the isotope effect exponent on $T_c$:
\begin{equation}
 \alpha=-\frac{\Delta T_c/T_c}{\Delta M/M},
 \label{eq:alpha-Tc}
\end{equation}
($M$ is the isotope mass and $\Delta M$ is the mass difference) and by further comparing it with the universal value $\alpha_{\rm BCS}\equiv0.5$ as is predicted within the framework of BCS theory of electron-phonon mediated
superconductivity.

In conventional phonon-mediated superconductors like simple metals, alloys, {\it etc.}  $\alpha$, typically, ranges from 0.2 to 0.5, (see Ref.~\onlinecite{Poole00} and references therein). The only exceptions are Ru and Zr exhibiting zero isotope effect and PdH(D) with $\alpha_{\rm H(D)}=-0.25$.\cite{IE-PdH1, IE-PdH2} The negative isotope effect of PdH(D) is explained, however, by the presence of strong lattice anharmonicty caused by the double-well potential in the proton (deuteron) bond distribution.\cite{Yussouff95}
%This was confirmed by neutron scattering data where the large zero point motion of H in comparison with that of Deuterium results in 20\% change of the lattice force constants.\cite{Rahman76}
%
A similar finding exists in organic superconductors where the H(D) isotope
effect changes sign as compared, {\it e.g.}, to $^{34}$S, $^{13}$C, and
$^{15}$N isotope replacements, (see Ref.~\onlinecite{Schlueter01}
and references therein). Again, an unusually strong anharmonic lattice dynamics
are attributed to this observation.\cite{Schlueter01,Whangbo97}
Recently the sine changed isotope effect was reported by Stucky {\it et al.}\cite{Stucky_arxiv_2016} for $n-$doped SrTiO$_3$ which supposed to be purely phonon-mediated superconductor.  It was observed that the substitution of the natural $^{16}$O atoms by the heavier isotope $^{18}$O causes a giant (of the order of 50\%) enhancement of $T_c$. Also the magnetic critical
field $H_{c2}$ is increased by a factor $\sim 2$. Such a strong impact on $T_c$ and $H_{c2}$, with a sign opposite to
conventional superconductors was assumed to be caused by strong coupling
to the ferroelectric soft modes of SrTiO$_3$.

The cuprate high-temperature superconductors (HTS) are characterized by a
vanishingly small but positive isotope effect exponent in optimally doped
compounds which increases in a monotonic way upon decreasing
doping.\cite{Batlogg87, Franck91, Franck94, Zech94, Zhao01, Khasanov_JPCM_2003, Khasanov_PRB_2003, Khasanov04, Khasanov04a, Khasanov06, Khasanov07, Khasanov08, Tallon05, IE_Bi2201-2212-2223, Khasanov08_IE-phase-diagram} For the optimally
doped cuprate HTS the smallest value of the  oxygen-isotope exponent
$\alpha_{\rm O}\simeq 0.02$ was obtained for YBa$_2$Cu$_3$O$_{7-\delta}$ and
Bi$_2$Sr$_2$Ca$_2$Cu$_3$O$_{10+\delta}$, while it reaches $\alpha_{\rm O}\simeq
0.25$ for
Bi$_2$Sr$_{1.6}$La$_{0.4}$CuO$_{6+\delta}$.\cite{Batlogg87,Franck91,Franck94,
Zech94,IE_Bi2201-2212-2223,Khasanov08_IE-phase-diagram} In addition, it was
demonstrated that in underdoped materials $\alpha_{\rm O}$ exceeds
substantially the BCS limit $\alpha_{\rm
BCS}\equiv0.5$.\cite{Franck91,Franck94,Zhao01,Khasanov08_IE-phase-diagram}

It is worth to emphasize here that the values of both, the oxygen and the copper
isotope exponents in cuprate HTS are {\it always} positive. Similar tendencies,
with the only few above mentioned exceptions, are also realized in a case of
conventional phonon-mediated superconductors.

Since the discovery of high-temperature superconductivity in Fe-based compounds few experiments to
study the isotope effect on $T_c$ in this new class of materials were performed. The current statement of isotope effect studies on Fe-based HTS remains, however, rather contradicting.
\cite{Liu09,Shirage09,Shirage10,Khasanov10_FeSe-isotope,Tsuge_PP_2012,Tsuge_JPCS_2014} Liu {\it et
al.}\cite{Liu09} and Khasanov {\it et al.}\cite{Khasanov10_FeSe-isotope} have
found a {\it positive} Fe isotope effect (Fe-IE) exponent $\alpha_{\rm Fe}$ for
Ba$_{0.6}$K$_{0.4}$Fe$_2$As$_2$ ($T_c\simeq 38$~K), SmFeAsO$_{0.85}$F$_{0.15}$ ($T_c\simeq 41$~K), and FeSe$_{1-x}$ ($T_c\simeq 8$~K)
with the corresponding values $\alpha_{\rm Fe}=0.34(3)$, 0.37(3), and 0.81(15),
respectively. Note that $\alpha_{\rm Fe}=0.81(15)$ for FeSe$_{1-x}$ exceeds
substantially  the universal BCS value $\alpha_{\rm BCS}\equiv0.5$. In the other studies Shirage {\it et al.} \cite{Shirage09,Shirage10} have
reported a {\it negative} $\alpha_{\rm Fe}=-0.18(3)$ and $\alpha_{\rm Fe}=-0.024(15)$ for
Ba$_{0.6}$K$_{0.4}$Fe$_2$As$_2$ ($T_c\simeq 38$~K) and
SmFeAsO$_{1-y}$ ($T_c\simeq 54$~K), respectively. The negative Fe-IE exponents were also reported recently  by Tsuge and co-workers for FeSe$_{0.35}$Te$_{0.65}$ ($T_c\simeq 15$~K, $\alpha_{\rm Fe}=-0.54$) and Ca$_{0.4}$Na$_{0.6}$Fe$_2$As$_2$ ($T_c\simeq 34$~K, $\alpha_{\rm Fe}=-0.19$).\cite{Tsuge_PP_2012,Tsuge_JPCS_2014}
Note that the sine changed isotope effect  is unlikely to stem from different pairing mechanisms going to be realized in different Fe-based superconductors. Especially, in the case of Ba$_{0.6}$K$_{0.4}$Fe$_2$As$_2$, when nominally {\it identical} samples exhibit once a positive\cite{Liu09} and next a negative isotope exponent.\cite{Shirage09}

The main purpose of the present study is to show that the very controversial results for Fe-IE on $T_c$ are caused by small structural changes occurring simultaneously with the Fe isotope exchange. We demonstrate that the
Fe-IE exponent on $T_c$ needs to be decomposed into the one related to the structural changes ($\alpha_{\rm Fe}^{\rm str}$)
and the genuine (intrinsic one, $\alpha_{\rm Fe}^{\rm int}$) to result in:
\begin{equation}
\alpha_{\rm Fe}=\alpha_{\rm Fe}^{\rm int}+\alpha_{\rm Fe}^{\rm str}.
 \label{eq:alpha-tot}
\end{equation}
The validity of such decomposition is further confirmed by the fact that $\alpha_{\rm Fe}^{\rm int}$ coincides
with the Fe-IE exponent on the characteristic phonon frequency $\omega_{\rm ph}$  as is observed in EXAFS and Raman experiments.\cite{Chu_SciRep_2013,Singh_AIP_2016} The value of $\alpha_{\rm Fe}^{\rm int}$ was found to be the same for representatives of various families of Fe-based superconductors and it stays in the range of $\alpha_{\rm Fe}^{\rm int} \sim0.3-0.4$ in good agreement with the theory prediction of Bussmann-Holder {\it et al.}\cite{Bussmann-Holder09}

The paper is organized as follows. In Section \ref{Sec:IE-FeSe} we demonstrate the influence of the Fe isotope exchange on the crystal structure. As an example, FeSe$_{1-x}$ Fe-based HTS with $T_c\simeq 8$~K is considered. It is shown that the substitution of
natural Fe (containing $\simeq92$\% of $^{56}$Fe) by its lighter $^{54}$Fe isotope leads not only to a shift of $T_c$, but affects also the structural parameters such as the lattice parameters $a$, $b$, and $c$, the lattice volume $V$, the distance between the Se atom and Fe plane, and the Se height $h_{\rm Se}$. Results presented in this section are adapted from Ref.~\onlinecite{Khasanov10_FeSe-isotope}. In Section~\ref{Sec:IE-General}
the currently available Fe isotope effect data on the
superconducting transition temperature $T_c$ for various Fe-based HTS were
reanalyzed by separating the measured Fe-IE exponent $\alpha_{\rm Fe}$ into a
structural and an intrinsic (unrelated to the structural changes) components. By taking structural corrections into account we infer that
the value of the genuine  Fe-IE exponent is close to $\alpha_{\rm Fe}^{\rm
int}\sim 0.3-0.4$. The results presented Section~\ref{Sec:IE-General} are partially adapted from Ref.~\onlinecite{Khasanov_PRB_2010}. The conclusions follow in Section~\ref{Sec:Conclusions}.

\section{${\rm \bf Fe}$ isotope effect on $T_c$ and the crystal structure of ${\rm \bf FeSe}_{1-x}$} \label{Sec:IE-FeSe}

In this section the $^{56}$Fe/${^{54}}$Fe isotope effects on the superconducting transition temperature and the crystal structure of the iron-chalcogenide superconductor FeSe$_{1-x}$ are described.

The preparation procedure for $^{56}$Fe/${^{54}}$Fe isotope submitted FeSe$_{1-x}$ powder samples is given in Ref.~\onlinecite{Khasanov10_FeSe-isotope}. 
The Fe-IE on the structural properties and the transition temperature $T_c$ were studied in neutron powder
diffraction (NPD) and magnetization experiments, respectively.

\subsection{Fe isotope effect on the crystal structure of FeSe$_{1-x}$}

\begin{figure}[htb]
 \begin{center}
\includegraphics[width=1.0\linewidth]{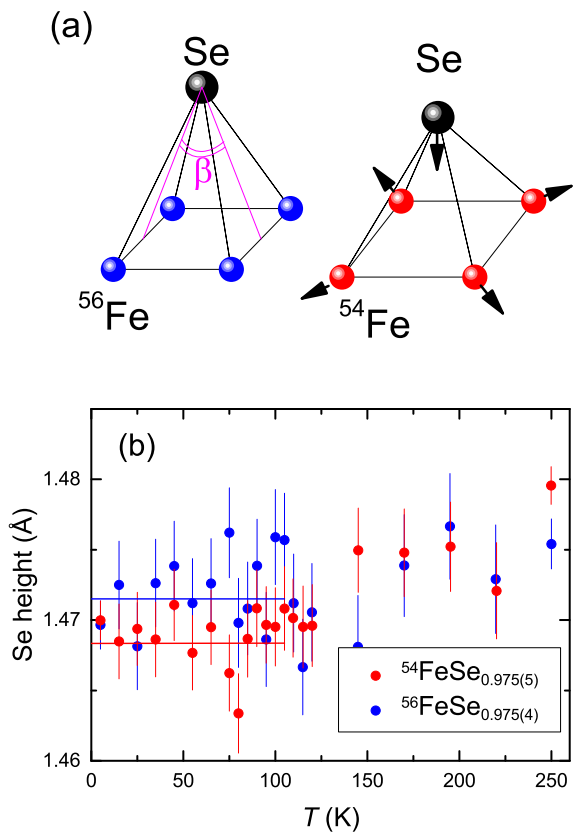}
 \end{center}
 \vspace{-0.5cm}
\caption{(a) The transformation of the Fe$_4$Se pyramid caused by
$^{56}$Fe/$^{54}$Fe isotope substitution in FeSe$_{1-x}$. The arrows show the direction of atom
displacements. (b) The temperature dependence of the distance between the Se atom and Fe
plane $h_{\rm Se}$. The straight lines correspond to averaging the data below 100~K. After Ref.~\onlinecite{Khasanov10_FeSe-isotope}. }
 \label{fig:NPD-lattice_parameters}
\end{figure}

The refined structural parameters at $T=250$~K and 5~K obtained in neutron powder
diffraction experiments are
summarized in Table~\ref{Table:NPD_IE-results}.
The results of NPD measurements imply that the substitution of $^{56}$Fe by $^{54}$Fe leads to a
small, but detectable {\it enhancement} of the lattice along the
crystallographic $a$ and $b$ directions and a {\it compression} of it along the
{\it c}-axis, resulting in a change of the shape of the Fe$_4$Se pyramid (see Fig.~\ref{fig:NPD-lattice_parameters}~a).
As is shown in Fig.~\ref{fig:NPD-lattice_parameters}~b, for temperatures below 100~K the
Se atom is located closer to the Fe plane in $^{54}$FeSe$_{1-x}$ than in
$^{56}$FeSe$_{1-x}$.

%%%%%%%%%%%%%%%%%%%%%%%%%%%%%%%%%%%%%%%%%%%%%%%%%%%%%%%
%
\begin{table}[htb]
\caption[~]{\label{Table:NPD_IE-results} Structural parameters of
$^{54}$FeSe$_{1-x}$ and $^{56}$FeSe$_{1-x}$ at $T=250$ and 5~K as obtained from  the refitment of neutron powder diffraction (NPD) data (after Ref.~\onlinecite{Khasanov10_FeSe-isotope}).
%The normalized chi-square parameter ($\chi^2$) determines the goodness of fit.
}
\begin{center}
% \vspace{-0.5cm}
\begin{tabular}{lcc|cc}\\
 \hline
 \hline
& \multicolumn{2}{c}{$T=250$~K}& \multicolumn{2}{c}{$T=5$~K}\\
&$^{54}$FeSe$_{1-x}$ & $^{56}$FeSe$_{1-x}$ & $^{54}$FeSe$_{1-x}$ & $^{56}$FeSe$_{1-x}$\\
%
%&&&&\\
%
\hline
Space group &  \multicolumn{2}{c|}{$P4/nmm$}& \multicolumn{2}{c}{$Cmma$}\\
Se content& 0.975(5) & 0.975(4)&\multicolumn{2}{c}{fixed to 0.975}\\
$a$({\AA})&\multirow{2}{*}{3.77036(3)}&\multirow{2}{*}{3.76988(5)}
&5.33523(10)&5.33426(10)\\
$b$({\AA})&&&5.30984(10)&5.30933(10)\\
$c$({\AA})&5.51619(9)&5.51637(9)&5.48683(9)&5.48787(9)\\
Volume (\AA$^3$)&156.883(3)&156.797(3)&155.438(5)&155.424(5)\\
$z-$Se&0.2319(2)&0.2326(0.3)&0.2321(2)&0.2322(3)\\
 \hline \hline \\

\end{tabular}
   \end{center}
\end{table}

It is
important to note that the observed Fe-IE's on the lattice parameters are
intrinsic and not just a consequence of slightly different samples. As shown in
Ref.~\onlinecite{Pomjakushina09}, various samples of $^{56}$FeSe$_{1-x}$  with $1-x
\simeq 0.98$ and $T_{c} \simeq 8.2$~K indeed exhibit the same lattice
parameters within the experimental error.

% Isotope effect on the lattice parameters

%%%%%%%%%%%%%%%%%%%%%%%%%%%%%%%%%%%%%%%%%%%%%%%%%%%
%
%

%%%%%%%%%%%%%%%%%%%%%%%%%%%%%%%%%%%%%%%%%%%%%%%%%%%%%%%

% Isotope effect on Tc

%%%%%%%%%%%%%%%%%%%%%%%%%%%%%%%%%%%%%%%%%%%%%%%%%%%%%
%
\begin{figure}[htb]
 \begin{center}
\includegraphics[width=1.0\linewidth]{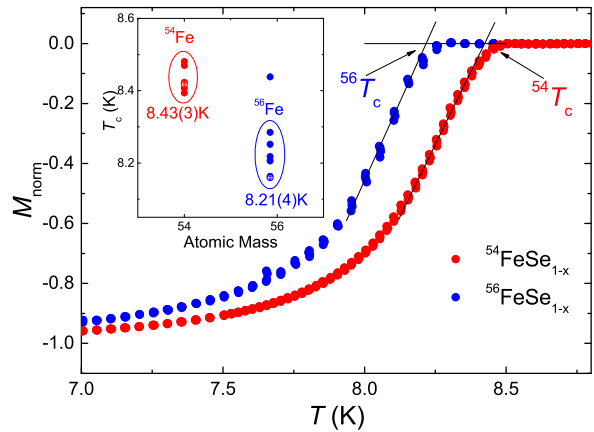}
 \end{center}
 \vspace{-0.5cm}
\caption{The normalized zero-field cooling magnetization curves $M_{norm}(T)$ for one pair of $^{54}$FeSe$_{1-x}$ and
$^{56}$FeSe$_{1-x}$ samples. The transition temperature $T_c$ is determined as
the intersection of the linearly extrapolated $M_{nomr}(T)$ curve in the
vicinity of $T_c$ with the $M=0$ line. The inset shows the superconducting transition temperature $T_c$ as a
function of Fe atomic mass for $^{54}$FeSe$_{1-x}$/$^{56}$FeSe$_{1-x}$ samples
studied in the present work. The open symbols correspond to the samples studied
by NPD experiments. The $T_c$'s fall into the regions marked by the colored
ovals with the corresponding mean values $^{54}\overline{T}_c=8.43(3)$~K and
$^{56}\overline{T}_c=8.21(4)$~K. After Ref.~\onlinecite{Khasanov10_FeSe-isotope}.}
 \label{fig:ZFC-DC}
\end{figure}
%
%%%%%%%%%%%%%%%%%%%%%%%%%%%%%%%%%%%%%%%%%%%%%%%%%%%%

\subsection{Fe isotope effect on $T_c$ of FeSe$_{1-x}$}

The Fe-IE on the transition temperature $T_c$ was studied by means of magnetization experiments. In order to avoid artifacts and systematic errors in the determination of the isotope shift of $T_{c}$ a {\it statistical} study were performed: {\it i.e.} the series of $^{54}$FeSe$_{1-x}$/${^{56}}$FeSe$_{1-x}$ samples synthesized exactly at the same way (the same thermal history, the same amount of Se in the initial composition) were investigated.

The magnetization experiments were conducted for six $^{54}$FeSe$_{1-x}$ and seven $^{56}$FeSe$_{1-x}$ samples,
respectively. Figure~\ref{fig:ZFC-DC} shows an example of zero-field cooled (ZFC) magnetization curves for a pair of $^{54}$FeSe$_{1-x}$/$^{56}$FeSe$_{1-x}$ samples ($M_{norm}$ was obtained after subtracting the small paramagnetic offset $M_{magn}$ measured at $T>T_c$ and further normalization of the obtained curve to the value at $T=2$~K, see Fig.~1 in Ref.~\onlinecite{Pomjakushina09} for details). The magnetization curve for $^{54}$FeSe$_{1-x}$ is shifted almost parallel to higher temperature, implying that $T_c$ of $^{54}$FeSe$_{1-x}$ is higher than that of $^{56}$FeSe$_{1-x}$. The resulting transition temperatures determined from the intercept of the linearly extrapolated $M_{norm}(T)$ curves with
the $M=0$ line for all samples investigated are summarized in the inset of  Fig.~\ref{fig:ZFC-DC}. The $T_c$'s for both sets of $^{54}$FeSe$_{1-x}$/$^{56}$FeSe$_{1-x}$ samples fall into two distinct regions: $8.39\leq {^{54}T_c} \leq 8.48$~K and $8.15\leq {^{56}T_c} \leq 8.31$~K, respectively. The corresponding mean values are: $^{54}\overline{T}_c=8.43(3)$~K and $^{56}\overline{T}_c=8.21(4)$~K. Note, that one out of the seven $^{56}$FeSe$_{1-x}$ samples had $T_c\simeq 8.44$~K which is by more than 5 standard deviations above the average calculated for the rest of the six samples. We have no explanation for this discrepancy, but decided to show this point for completeness of the data collected.

The value of the Fe-IE exponent $\alpha_{\rm Fe}$ determined from the data presented in Fig.~\ref{fig:ZFC-DC} by means of Eq.~\ref{eq:alpha-Tc} leads to $\alpha_{\rm Fe}= 0.81(15)$.

\subsection{The structural and the intrinsic contributions to the Fe-IE in FeSe$_{1-x}$}

The Fe-IE exponent $\alpha_{\rm Fe}= 0.81(15)$ obtained for FeSe$_{1-x}$ iron-chalcogenide superconductor is larger than
the BCS value $\alpha_{\rm BCS}\equiv0.5$ as well as more than twice as
large as $\alpha_{\rm Fe}\simeq0.35$ reported for
SmFeAsO$_{0.85}$F$_{0.15}$ and Ba$_{0.6}$K$_{0.4}$Fe$_2$As$_2$.\cite{Liu09}
We want to emphasize, however, that our structural refined NPD data suggest that part of the large Fe-IE exponent
$\alpha_{\rm Fe}= 0.81(15)$ may result from the tiny structural
changes occurring due to $^{54}$Fe/$^{56}$Fe substitution.

An estimate of the structural contribution to the Fe-IE exponent in FeSe$_{1-x}$ can be made
based on the observed proportionality between $T_{c}$ and the crystal lattice constant 
$a$ for the 11 family FeSe$_{1-y}$Te$_{y}$.\cite{Horigane09,Mizuguchi09} Assuming the relation $T_{c}$ vs. {\it a}  also holds for FeSe$_{1-x}$ one obtains from the data presented in Ref.~\onlinecite{Mizuguchi09} for $y\leq 0.5$ the relation
$\Delta T_c^a/(\Delta a/a) \approx 6$~K/\%.
With $(\Delta a +\Delta b)/(a+b)=0.0195(14)$\% (see Table~\ref{Table:NPD_IE-results} and Ref.~\onlinecite{Khasanov10_FeSe-isotope}) this results in a {\em structural}
shift of $T_c$ of $\Delta T_c^{\rm str}\approx 0.1$~K. Taking this correction into account yields a rough estimate of the structural and the intrinsic Fe-IE exponents of $\alpha_{\rm Fe}^{\rm str} \simeq \alpha_{\rm Fe}^{\rm int} \simeq 0.4$. Note that the value of $\alpha_{\rm Fe}^{\rm int}\simeq 0.4$ is comparable with
$\alpha_{\rm Fe}\simeq 0.35$ reported for SmFeAsO$_{0.85}$F$_{0.15}$ and Ba$_{0.6}$K$_{0.4}$Fe$_2$As$_2$.\cite{Liu09}

We want to stress, that the ''structural`` contribution into the Fe-IE exponent in FeSe$_{1-x}$ consists of both: a positive and negative parts. Indeed, in FeSe$_{1-x}$ a decrease of the Se height caused by
compression of the Fe$_4$Se pyramid (see Fig.~\ref{fig:NPD-lattice_parameters}~b) leads to an increase of $T_c$
by $\Delta T_c^{h_{\rm Se}}/(\Delta h_{\rm Se}/h_{\rm Se})\simeq
3.4$~K/\% \cite{Mizuguchi10,Margadonna09}. In contrast, an
increase of the Se(Te)-Fe-Se(Te) angle in the FeSe$_{1-y}$Te$_{y}$
family (angle $\beta$ in our notation, see Fig.~\ref{fig:NPD-lattice_parameters}~a) results for
$y\leq0.5$ in a decrease of $T_c$ by $\Delta T_c^\beta/(\Delta
\beta/\beta)\simeq 2.9$~K/\% \cite{Horigane09}. Based on the
structural data presented in Fig.~\ref{fig:NPD-lattice_parameters} and Table~\ref{Table:NPD_IE-results}
one obtains $\Delta h_{\rm Se}/h_{\rm Se}=0.22(8)$\% and
$\Delta\beta/\beta=-0.13(4)$\%, leading to $\Delta T_c^{h_{\rm
Se}}=0.7(3)$~K and $\Delta T_c^\beta=-0.4(2)$~K (in this estimate
the values of $h_{\rm Se}$
were averaged over the temperature regions denoted as solid lines
in Fig.~\ref{fig:NPD-lattice_parameters}~b).
Bearing in mind that all Fe-based HTS are similarly sensitive to
structural changes as FeSe$_{1-x}$ (see, {\it e.g.},
Refs.~\onlinecite{Horigane09,Mizuguchi10,Mizuguchi09,Zhao08}) we conclude that depending on the particular structural changes caused by the Fe isotope substitution, the shift of the transition temperature $T_c$ could be either {\it positive}, {\it negative} or may stay at {\it zero}. Which particular case is going to be realized would require precise studies of the structural properties for each individual isotope experiment.
%%%%%%%

\section{Genuine (Intrinsic) and structural isotope effects in ${\rm \bf Fe}$-based HTS} \label{Sec:IE-General}

\begin{figure}[htb]
%\centering
\includegraphics[width=1.0\linewidth]{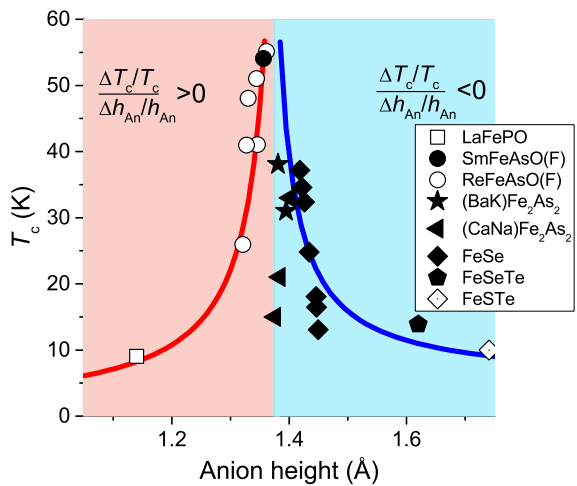}
 \vspace{-0.5cm}
\caption{Dependence of the superconducting transition
temperature $T_c$ on the height of the anion atom ($h_{\rm An}$, An=As, Se, P)
for varios families of Fe-based HTS, after Mizuguchi {\it et
al.}\cite{Mizuguchi10} The closed symbols represent the samples which are
relevant for the present study. The lines are guided for the eye. The red(blue)
area represents part of $T_c$ vs. $h_{\rm An}$ diagram where $T_c$
increases(decreases) with increasing $h_{\rm An}$. }
 \label{fig:Pnictogen-height}
\end{figure}

The observation of tiny but not negligible isotope effect on the crystal structure observed in FeSe$_{1-x}$ Fe-based HTS reported in the previous section requires to separate the Fe-IE exponent on $T_c$  into the genuine (intrinsic)  and the structural components (see Eq.~\ref{eq:alpha-tot}). In fact, the superconductivity in all Fe-based HTS is intimately related to small structural
changes as reported in various works. As an example the strong nonlinear dependence of the superconducting transition temperature on the anion atom height ($h_{\rm An}$, An=As, P, or Se) with a sharp maximum of $T_c$ at $h_{\rm An}\simeq1.38$~\AA, was reported by Mizuguchi {\it et al.}\cite{Mizuguchi10} (see Fig.~\ref{fig:Pnictogen-height}).

\begin{table*}[htb]
\caption[~]{\label{Table:IE-results}  Summary of Fe isotope effect studies for
SmFeAsO$_{0.85}$F$_{0.15}$ and Ba$_{0.6}$K$_{0.4}$Fe$_2$As$_2$ Ref.~\onlinecite{Liu09}, Ba$_{0.6}$K$_{0.4}$Fe$_2$As$_2$
Ref.~\onlinecite{Shirage09}, SmFeAsO$_{1-x}$ Ref.~\onlinecite{Shirage10}, FeSe$_{1-x}$ Ref.~\onlinecite{Khasanov10_FeSe-isotope}, FeSe$_{0.35}$Te$_{0.65}$ Ref.~\onlinecite{Tsuge_PP_2012} and Ca$_{0.4}$Na$_{0.6}$Fe$_2$As$_2$ Ref.~\onlinecite{Tsuge_JPCS_2014}.
The parameters are: $T_c$ -- superconducting transition tempreature for the
sample with the natural Fe isotope ($^{\rm nat}$Fe); $\alpha_{\rm Fe}$ -- Fe
isotope effect exponent; c -- the {\it c}-axis lattice constant for the sample with
the lighter ($^{\rm light}$Fe) and the heavier ($^{\rm heavy}$Fe) Fe isotope;
$\Delta c/c$ -- the relative shift of the {\it c}-axis constant caused by the Fe
isotope substitution; $\alpha_{\rm Fe}^{\rm str}$ and $\alpha_{\rm Fe}^{\rm
int}$ -- the structural and the intrinsic contributions to $\alpha_{\rm Fe}$.
}
\begin{center}
% \vspace{-0.5cm}
\begin{tabular}{ccccccccccc}\\
 \hline
 \hline
%
%
%&& \multicolumn{3}{c}{Light Fe isotope}&&& \multicolumn{3}{c}{Heavy Fe isotope}&&\\
%
%
Sample&Reference&$T_c(^{\rm nat}$Fe)&$\alpha_{\rm Fe}$&{\it c}-axis($^{\rm
light}$Fe)&{\it c}-axis($^{\rm heavy}$Fe)&$\Delta {\rm c}/{\rm c}$
&$\alpha_{\rm Fe}^{\rm str}$&$\alpha_{\rm Fe}^{\rm int}$\\
&&(K)&&(\AA)&(\AA)&&&\\
\hline
Ba$_{0.6}$K$_{0.4}$Fe$_2$As$_2$&Ref.~\onlinecite{Liu09}&37.30(2) &0.37(3) &13.289(7)&13.288(7)&$\sim0$&$\sim0$&$\sim0.35$ \\
Ba$_{0.6}$K$_{0.4}$Fe$_2$As$_2$&Ref.~\onlinecite{Shirage09}&37.78(2) &$-0.18(3)$ &13.313(1)&13.310(1)&$<0$&$\sim-0.5$&--\\
SmFeAsO$_{0.85}$F$_{0.15}$&Ref.~\onlinecite{Liu09}&41.40(2) &0.34(3) &8.490(2)&8.491(2)&$\sim0$&$\sim0$&$\sim0.35$ \\
SmFeAsO$_{1-y}$&Ref.~\onlinecite{Shirage10}&54.02(13) &$-0.024(15)$ &8.4428(8)&8.4440(8)&$\gtrsim0$&$\sim-0.35$& --\\
FeSe$_{1-x}$ & Ref.~\onlinecite{Khasanov10_FeSe-isotope}&8.21(4) &0.81(15) &5.48683(9)&5.48787(9)&$>0$&$\simeq0.4$&$\simeq0.4$ \\
FeSe$_{0.35}$Te$_{0.65}$&Ref.~\onlinecite{Tsuge_PP_2012}& 13.10(5)& -0.54& -- & --  & -- & -- &--\\
Ca$_{0.4}$Na$_{0.6}$Fe$_2$As$_2$& Ref.~\onlinecite{Tsuge_JPCS_2014} &34.3(2)&-0.19&  12.19& 12.19 & -- & -- &-- \\

 \hline \hline \\

\end{tabular}
   \end{center}
\end{table*}

\subsection{Summary of Fe-IE measurements on the transition temperature $T_c$} \label{Subsec:IE-FeSe}

\begin{figure}[htb]
%\centering
\includegraphics[width=1.0\linewidth]{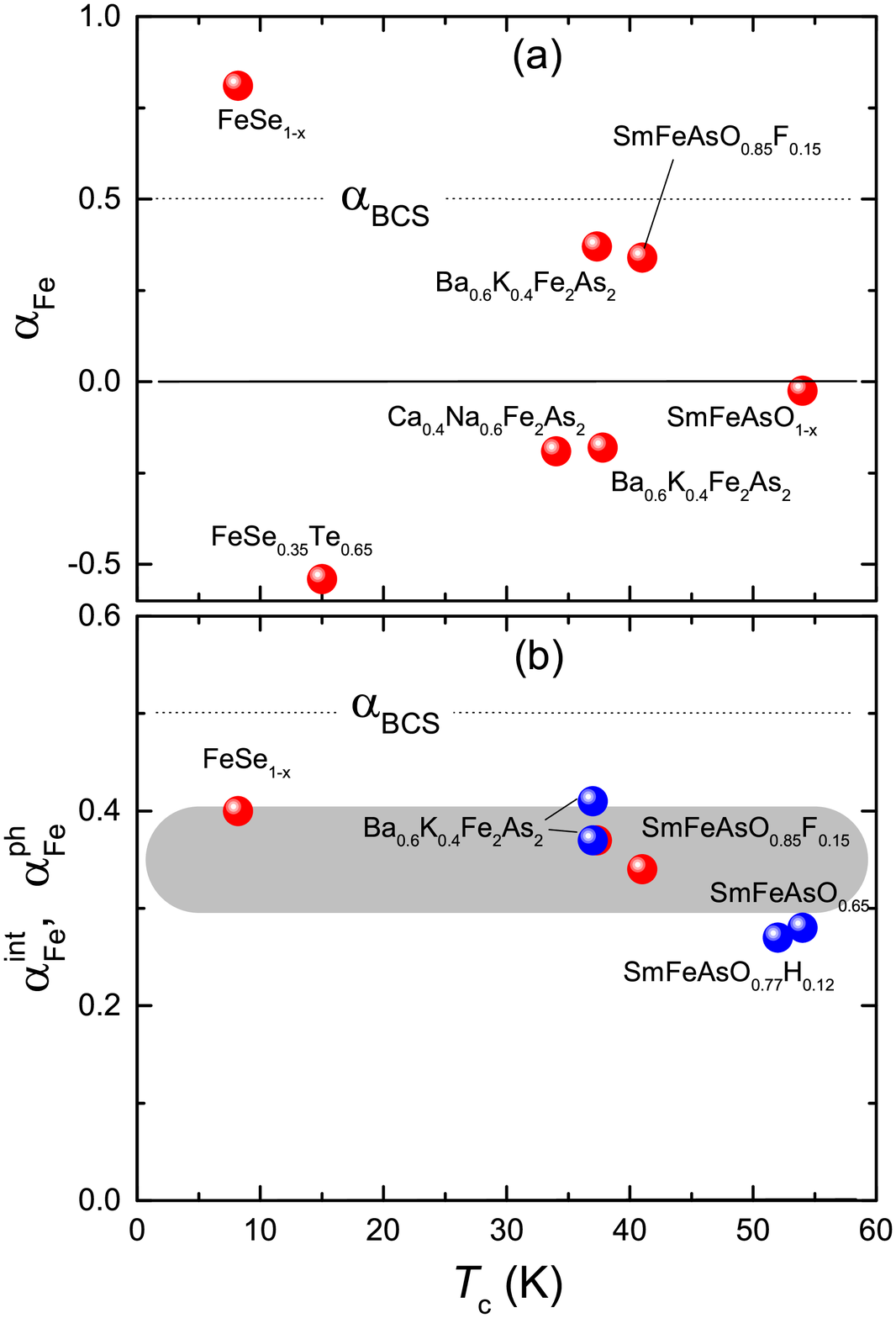}
 \vspace{-0.5cm}
\caption{(a) Fe isotope effect exponent $\alpha_{\rm Fe}$ as a
function of the superconducting transition temperature $T_c$ for the samples
considered in the present study: Ba$_{0.6}$K$_{0.4}$Fe$_2$As$_2$ and
SmFeAsO$_{0.85}$F$_{0.15}$ Ref.~\onlinecite{Liu09},
Ba$_{0.6}$K$_{0.4}$Fe$_2$As$_2$ Ref.~\onlinecite{Shirage09},
SmFeAsO$_{1-x}$ Ref.~\onlinecite{Shirage10}, FeSe$_{1-x}$
Ref.~\onlinecite{Khasanov10_FeSe-isotope}, FeSe$_{0.35}$Te$_{0.65}$ Ref.~\onlinecite{Tsuge_PP_2012} and Ca$_{0.4}$Na$_{0.6}$Fe$_2$As$_2$ Ref.~\onlinecite{Tsuge_JPCS_2014}. $\alpha_{\rm
BCS}\equiv0.5$ denotes the BCS value for electron-phonon mediated superconductivity. (b) The ``intrinsic'' $\alpha_{\rm
Fe}^{\rm int}$ (red circles) and the phonon $\alpha_{\rm
Fe}^{\rm ph}$ (blue circles) Fe isotope effect exponents as they obtained in the present study and the EXAFS\cite{Chu_SciRep_2013} and Raman\cite{Singh_AIP_2016} experiments, respectively. The grey area denotes the region of $0.3\leq \alpha \leq 0.4$. See text for details. }
 \label{fig:Isotope-exponent}
\end{figure}

The results of Fe-IE measurements of $T_c$ for various Fe-based HTS's obtained up to date are summarized in Table~\ref{Table:IE-results} and Fig.~\ref{fig:Isotope-exponent}~a. In order to account for the structural changes caused by the Fe isotope exchange we have also included in Table~\ref{Table:IE-results} the {\it c}-axis lattice constants as they measured for the pairs of Fe isotope substituted samples. The choice of the {\it c}-axis lattice
constant as the relevant quantity in deriving the structural isotope effect
might appear to be rather arbitrary since $T_c$ is influenced by all structural
details, namely tetrahedral angle, {\it a}-axis lattice constant, internal bond
lengths, {\it etc}. However, the {\it c}-axis lattice constant provides a very
sensitive probe since its compression(expansion) is directly accompanied by the
corresponding variation of the distance from the Fe-planes to the above(below)
lying anions which, in turn, is a well characterized property for many Fe-based
compounds.\cite{Mizuguchi10} A survey of the literature shows that the
proportionality between the anion atom height and the {\it c}-axis lattice constant
indeed holds for various families of
Fe-based HTS considered in the present
study.\cite{Margadonna09,Rotter08,McQueen09}

From Table ~\ref{Table:IE-results} it follows that in
Ba$_{0.6}$K$_{0.4}$Fe$_2$As$_2$ and SmFeAsO$_{0.85}$F$_{0.15}$ studied in Ref.~\onlinecite{Liu09} the
{\it c}-axis constants are the same within the experimental error for both
isotopically substituted sets of the samples. In FeSe$_{1-x}$
\cite{Khasanov10_FeSe-isotope} the {\it c}-axis constant is larger, while in
Ba$_{0.6}$K$_{0.4}$Fe$_2$As$_2$ \cite{Shirage09} it is smaller for the sample
with the heavier Fe isotope. In SmFeAsO$_{1-y}$, studied by Shirage {\it et
al.},\cite{Shirage10} both {\it c}-axis lattice constants seem to coincide within the
experimental resolution. However, since the difference between them is 1.5
times larger than one standard deviation, it is conceivable to attribute an
increase in the {\it c}-axis lattice constant in SmFeAsO$_{1-y}$ with the heavier Fe
isotope. It should also be noted here that the accuracy in lattice constants determination for Ca$_{0.4}$Na$_{0.6}$Fe$_2$As$_2$ from Ref.~\onlinecite{Tsuge_JPCS_2014} is from one to two orders of magnitude worse than it is reported for other isotope substituted samples, and no structural data were reported for FeSe$_{0.35}$Te$_{0.65}$ in Ref.~\onlinecite{Tsuge_PP_2012}.

The use of the empirical $T_c$ vs. $h_{\rm An}$ relation from
Ref.~\onlinecite{Mizuguchi10} combined with the intrinsic relation of the
proportionality between the {\it c}-axis constant and the anion atom height
($\Delta{\rm c}\propto\Delta h_{\rm An}$, see
Refs.~\onlinecite{Margadonna09,Rotter08,McQueen09}) enables us to determine the
sign of the structurally related shift of $T_c$  induced by isotopic exchange.
By 
%defining the shift of a given quantity $X$ as $\Delta X/X=(\;^{^{\rm light}{\rm Fe}}X-\;^{^{\rm heavy}{\rm Fe}}X)/\;^{^{\rm heavy}{\rm Fe}}X$ and
following Mizuguchi {\it et al.}\cite{Mizuguchi10}  (see also
Fig.~\ref{fig:Pnictogen-height}) the sign of $(\Delta T_c/T_c)/(\Delta h_{\rm
An}/h_{\rm An})$ is positive for SmFeAsO(F) as well as for various Fe-based HTS
belonging to ReFeAsO(F) family (Re=Nd, Ce, La) and negative for
(BaK)Fe$_2$As$_2$, (CaNa)Fe$_2$As$_2$, Fe(SeTe), and FeSe$_{1-x}$. Consequently the change of the {\it c}-axis
constant caused by Fe isotope substitution as presented in
Table~\ref{Table:IE-results} results in an additional structurally related
shift of $T_c$ being positive for FeSe$_{1-x}$,\cite{Khasanov10_FeSe-isotope}
negative for Ba$_{0.6}$K$_{0.4}$Fe$_2$As$_2$ and
SmFeAsO$_{1-y}$,\cite{Shirage09,Shirage10} and close to 0 for
Ba$_{0.6}$K$_{0.4}$Fe$_2$As$_2$ and SmFeAsO$_{0.85}$F$_{0.15}$.\cite{Liu09} 
Note that for the lack of pure(absent) structural data, such estimate is not possible to be made for Ca$_{0.4}$Na$_{0.6}$Fe$_2$As$_2$ and FeSe$_{0.35}$Te$_{0.65}$ from Ref.~\onlinecite{Tsuge_PP_2012,Tsuge_JPCS_2014}.

We want to emphasize that the above mentioned discussion allows only to determine the sign of
the structurally related isotope effect but not its absolute value. The reasons
are the following. (i) The relative change of the {\it c}-axis constant is
proportional, but not identical to the one of $h_{\rm An}$. As an example,
$^{56}$Fe to $^{54}$Fe isotope substitution in FeSe$_{1-x}$ leads to an
increase of the {\it c}-axis constant by approximately 0.02\%, while the change of
the Se height amounts to ten times higher value $\simeq0.22$\% (see Table~\ref{Table:NPD_IE-results}, Fig.~\ref{fig:NPD-lattice_parameters}~b, and
Ref.~\onlinecite{Khasanov10_FeSe-isotope}). (ii) The height of the anion atom
is clearly not the only parameter which is crucial for $T_c$ of Fe-based HTS as
already mentioned above.
%However, the lack of a consistent structural characterization limits this study to a single parameter which was emphasized to be of the uppermost relevance to $T_c$.

The analysis of the structural data together with the dependence of $T_c$ on the {\it a}-axis lattice constant  in FeSe$_{1-x}$, as is presented in Section~\ref{Sec:IE-FeSe}, was allowing to extract the
``structural'' Fe isotope effect exponent $\alpha_{\rm Fe}^{\rm str}\simeq 0.4$
for $^{56}$Fe to $^{54}$Fe substituted FeSe$_{1-x}$
samples. The absence of precise structural data for the samples studied in Refs.~\onlinecite{Liu09,Shirage09,Shirage10,Tsuge_PP_2012,Tsuge_JPCS_2014}
complicates their analysis.
However, a zero, within the experimental accuracy, Fe isotope shift of the
{\it c}-axis lattice constant for Ba$_{0.6}$K$_{0.4}$Fe$_2$As$_2$ as reported by Liu
{\it et al.}\cite{Liu09} is a clear indication that no structural effect is
present for this particular set of the samples. Consequently, the negative
isotope effect exponent $\alpha_{\rm Fe}\simeq-0.18$ obtained for nominally
identically doped Ba$_{0.6}$K$_{0.4}$Fe$_2$As$_2$ by Shirage {\it et
al.}\cite{Shirage09} stems from summing both effects, {\it i.e.},
$-0.18(\alpha_{\rm Fe})=0.35(\alpha_{\rm Fe}^{\rm int} )-0.52(\alpha_{\rm
Fe}^{\rm str} )$, see Eq.~(\ref{eq:alpha-tot}).
The similar analysis by comparing SmFeAsO(F) Fe-IE exponents from Refs.~\onlinecite{Liu09} and \onlinecite{Shirage10} results in $\alpha_{\rm Fe}^{\rm str}\simeq -0.35$ for SmFeAsO$_{1-y}$ from Ref.~\onlinecite{Shirage10}. It is worth to note, however, that the SmFeAsO(F) samples studied in Refs.~\onlinecite{Liu09} and \onlinecite{Shirage10} have different doping levels (different $T_c$'s, see Table~\ref{Table:IE-results} and Fig.~\ref{fig:Isotope-exponent}~a).

\subsection{The genuine Fe-IE exponent via phonon frequency measurements}

The BCS expression for the superconducting transition temperature $T_c$ relates it to the characteristic phonon frequency $\omega_{\rm ph}$ and the coupling constant $\lambda$ via:\cite{Tinkham_75}
\begin{equation}
T_c\propto \omega_{\rm ph} \exp(-1/\lambda).
 \label{eq:BCS-Tc}
\end{equation}
The consequences of this equation are two fold.
First of all, since the coupling constant $\lambda$ is independent of the ion mass $M$ (see {\it e.g.} Ref.~\onlinecite{Alexandrov_book_2003}) and the characteristic phonon frequency $\omega_{\rm ph}$ is proportional to $1/\sqrt{M}$, as a frequency of any harmonic oscillator, $\alpha_{\rm BCS}$ becomes equal exactly to 0.5. Note that the Coulomb repulsion and the anharmonicity of phonons, which are not considered in Eq.~\ref{eq:BCS-Tc}, would lead to smaller $\alpha$ values.\cite{Alexandrov_book_2003}
Second, due to direct proportionality between $T_c$ and $\omega_{\rm ph}$ the isotope exponents on both these quantities are equal to the each other:
\begin{equation}
 \alpha=-\frac{\Delta T_c/T_c}{\Delta M/M}\equiv -\frac{\Delta \omega_{\rm ph}/\omega_{\rm ph}}{\Delta M/M}.
 \label{eq:alpha_Tc-ph}
\end{equation}

Currently we are aware of two experimental works where the Fe-IE on $\omega_{\rm ph}$ was studied by means of EXAFS and Raman techniques on (BaK)Fe$_2$As$_2$, SmFeAsO$_{0.65}$, and SmFeAs$_{0.77}$H$_{0.12}$ Fe-based HTS.\cite{Chu_SciRep_2013,Singh_AIP_2016} The corresponding $\alpha^{\rm ph}_{\rm Fe}$'s are shown in Fig.~\ref{fig:Isotope-exponent}~b together with presently obtained $\alpha^{\rm int}_{\rm Fe}$'s (see Section~\ref{Subsec:IE-FeSe} and Table~\ref{Table:IE-results}). Remarkably enough both sets of Fe-IE exponents stay quite close to the each other with $\alpha^{\rm int}_{\rm Fe}\simeq\alpha^{\rm ph}_{\rm Fe} \simeq 0.35\pm 0.07$.

\subsection{Isotope effect within the multiple gap scenario of ${\rm \bf Fe}$-based HTS}

Bussmann-Holder {\it et al.}\cite{Bussmann-Holder09} investigated a
multiple gap scenario of superconductivity in Fe-based HTS with the aim to
search for possible sources of the isotope effect on $T_c$. Typical phonon-mediated scenarios were contrasted to polaronic effects and found to have very
different impacts on the isotope effect. While phonon-mediated
superconductivity slightly suppresses the isotope effect as compared to the BCS
value $\alpha_{\rm BCS}\equiv0.5$, polaronic effects can largely enhance it.
The scenario of electron-phonon mediated superconductivity within the dominant
gap channel predicts a $T_c$ independent isotope effect with the $\alpha$ value
being slightly smaller than 0.5. This  agrees rather well with that observed
for FeSe$_{1-x}$,\cite{Khasanov10_FeSe-isotope}
(BaK)Fe$_2$As$_2$,\cite{Liu09,Shirage09,Chu_SciRep_2013} 
SmFeAsO$_{0.85}$F$_{0.15}$,\cite{Liu09}  SmFeAsO$_{0.65}$,\cite{Singh_AIP_2016} and SmFeAs$_{0.77}$H$_{0.12}$.\cite{Singh_AIP_2016} Indeed, for these particular samples,
which are belong to 3 different families of Fe-based HTS and have $T_c$'s ranging
from 8 to 54~K, the ``intrinsic'' Fe isotope exponent is almost constant with
$\alpha_{\rm Fe}^{\rm int}\sim0.3-0.4$, see Table~\ref{Table:IE-results} and
Fig.~\ref{fig:Isotope-exponent}~b.

\section{Conclusions} \label{Sec:Conclusions}

The purpose of the present study was two-fold.
First, by presenting results obtained on FeSe$_{1-x}$ iron-chalcogenide superconductor it was demonstrated that the tiny changes of the
structural parameters caused by Fe isotope substitution are clearly contributing to the resulting Fe-IE exponent on the transition temperature $T_c$.
Depending on the particular changes caused by the Fe isotope substitution, the ''structurally`` related shift of the transition temperature $T_c$ could be either positive, negative or stays at zero.
Such effects may help to clarify or even be the origin of the previously reported
controversial results.\cite{Liu09,Shirage09,Shirage10, Khasanov10_FeSe-isotope, Tsuge_PP_2012,Tsuge_JPCS_2014}
%However, more detailed and systematic structural investigations on Fe isotope substituted samples are required in order to draw definite conclusions.

Second, the currently available Fe isotope effect data on the
superconducting transition temperature $T_c$ for various Fe-based HTS were
analyzed by separating the measured Fe-IE exponent $\alpha_{\rm Fe}$ into a
structural ($\alpha_{\rm Fe}^{\rm str}$) and an intrinsic (unrelated to the structural changes, $\alpha_{\rm Fe}^{\rm int}$) components. The validity of decomposition the Fe-IE exponent $\alpha_{\rm Fe}$ was further confirmed by the fact that $\alpha_{\rm Fe}^{\rm int}$ almost coincides
with the Fe-IE exponent on the characteristic phonon frequencies as observed in recent EXAFS and Raman experiments.\cite{Chu_SciRep_2013,Singh_AIP_2016}
We infer that the value of the genuine  Fe-IE exponent is close to $\alpha_{\rm Fe}^{\rm
int}\sim 0.3-0.4$ for compounds having $T_c$ ranging from 8 to 54~K and belonging to at least three different families
of Fe-based HTS.

\section{Acknowledgments}

The author acknowledges the broad and successful collaboration with people without whom this work could not be performed: Markus Bendele, Hugo Keller, Annette Bussmann-Holder, Kazimierz Conder, Ekaterina Pomjakushina, and Volodya Pomjakushin. The special thank goes to Karl Alex M\"{u}ller and Hugo Keller who were initiated my interest to the isotope effect studies.

\end{document}